\documentclass[twocolumn,showpacs,preprintnumbers,amsmath,amssymb]{revtex4}

\usepackage{graphicx}
\usepackage{dcolumn}
\usepackage{bm}

\begin{document}

\preprint{APS/123-QED}

\title{Precision spectroscopy and density-dependent frequency shifts in ultracold Sr}

\author{Tetsuya Ido, Thomas H. Loftus, Martin M. Boyd, Andrew D. Ludlow, Kevin W. Holman, and Jun Ye}

\affiliation{%
JILA, National Institute of Standards and Technology and University of Colorado\\
Boulder, Colorado  80309-0440, U.S.A.\\
}%

\date{\today}
\begin{abstract}

By varying the density of an ultracold $^{88}$Sr sample from
$10^9$ cm$^{-3}$ to $> 10^{12}$ cm$^{-3}$, we make the first
definitive measurement of the density-related frequency shift and
linewidth broadening of the $^1S_0$ - $^3P_1$ optical clock
transition in an alkaline earth system. In addition, we report the
most accurate measurement to date of the $^{88}$Sr $^1S_0 - ^3P_1$
optical clock transition frequency. Including a detailed analysis
of systematic errors, the frequency is ($434\; 829 \;121\; 312\;
334 \pm 20_{stat} \pm 33_{sys}$) Hz.
\end{abstract}

\maketitle

Neutral-atom-based optical frequency standards are becoming
serious contenders for the realization of highly stable and
accurate optical atomic clocks \cite{hollberg01, wilpers02,
katori03}. Progress in this field is rapid, with significant
advances in experimental techniques from several related fields
including laser cooling to ultracold temperatures \cite{katori99,
binnewies01, mukaiyama03, loftus04}, trapping configurations
suitable for highly accurate clocks \cite{katori03}, the
development of high quality optical local oscillators
\cite{rafac00}, and precise frequency measurement and
distribution enabled by femtosecond optical combs \cite{diddams01,
ye01, cundiff02}.

Several systems have been proposed for neutral-atom-based optical
clocks, each emphasizing narrow intercombination transitions in
alkaline earth \cite{hollberg01, wilpers02, ruschewitz98, ido03,
xu03} and Yb atoms \cite{park03, porsev04}. High resolution
spectroscopic techniques employed to date include recoil-free
absorption inside 1-D optical lattices \cite{ido03}, free space
Ramsey interrogation and atom interferometry \cite{trebst01}, and
free space saturated absorption \cite{oates04}. The fermionic
strontium isotope, $^{87}$Sr, offers a weakly allowed $^1S_0$ -
$^3P_0$ optical clock transition with a potential resonance
quality factor exceeding $10^{15}$ and reduced polarization
dependence and collisional shifts of the clock frequency when the
atoms are confined inside magic-wavelength optical lattices
\cite{katori03, courtillot03, xu03b}. However, the large nuclear
spin ($I = 9/2$) brings complexity in state preparation and field
control. Recently new schemes have emerged that take advantage of
the most abundant spin-zero isotope ($^{88}$Sr) by engineering a
$^1S_0$ - $^3P_0$ clock transition with a perfect scalar nature
\cite{hong04}.

In this Letter, we present precision spectroscopy of ultracold
$^{88}$Sr in free space. With reference to the Cs primary standard
and with a detailed investigation of systematic frequency shifts,
we determine the absolute frequency of the $^1S_0 - ^3P_1$ clock
transition with a statistical (systematic) uncertainty of $20(33)$
Hz. This level of measurement precision represents an improvement
of more than 200 times over recent studies performed with thermal
atomic beams \cite{ferrari03}. Collision-related frequency shifts
can significantly impact both the stability and accuracy of
microwave and optical frequency standards. These effects have
been studied in detail for both Cs and Rb microwave clocks
\cite{fertig00, sortais00, leo01}. For optical frequency
standards based on intercombination transitions in alkaline earth
atoms, it is expected that the dependence of the fractional
frequency shift on the atomic density is reduced by more than
three orders of magnitude in comparison to Cs clocks
\cite{julienne04}. Previous efforts to measure this effect in Ca,
however, were hampered by low sample densities $\sim 10^9$
cm$^{-3}$, resulting in measured shift coefficients with errors
exceeding their associated absolute values \cite{hollberg04}. To
overcome this limitation, we have performed absolute frequency
measurements with $\sim$1 $\mu$K $^{88}$Sr atoms for densities
spanning three orders of magnitude, from $10^9$ cm$^{-3}$ to
$10^{12}$ cm$^{-3}$. For the first time we report definitive
density-related line center shifts and spectral broadening of an
optical clock transition in ultracold atoms. We find the
density-dependent fractional frequency shift is 250 times smaller
than that of Cs.

\begin{figure}
\resizebox{8.5cm}{!}{
\includegraphics{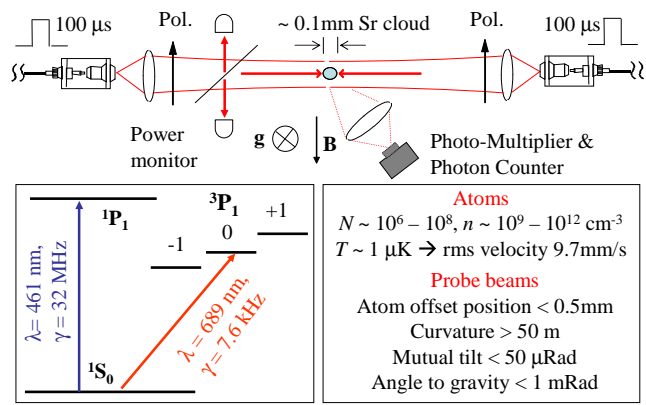}}
\caption{\label{Fig1} (color online) Top view of the ultracold
$^{88}$Sr spectrometer. The two counter-propagating beams are
collimated and overlapped. The beam intensities are balanced via
atomic fluorescence signals. Light polarization is parallel to
the bias magnetic field. A simplified $^{88}$Sr level diagram and
relevant experimental parameters are given in the bottom two
panels.}
\end{figure}

Figure 1 shows a simplified view of the ultracold $^{88}$Sr
spectrometer. Components pertaining to atom cooling and trapping
are described elsewhere \cite{loftus04, loftus04b}. Measurements
reported here utilize $10^6$ to $10^8$ atoms cooled to $\sim$ 1.3
$\mu$K. Spectroscopic probing of the $^1S_0$ - $^3P_1$ clock
transition is performed in free space after the trapping laser
beams and the quadrupole magnetic field are extinguished. The
atomic density $n$ during spectroscopic probing is varied by
changing either the initial atom number in the trap or the
free-flight time after the atoms are released to free space. The
typical free-flight times range from 5 to 20 ms. Spatial
densities are determined from calibrated fluorescence images. The
probe laser linewidth is $<$100 Hz. The laser's absolute
frequency is measured continuously via a femtosecond optical comb
that is locked to a Cs-referenced hydrogen maser via a fiber
optic link to NIST \cite{ye03}. The probe laser is split into two
counter-propagating beams, each coupled respectively into a
single-mode, polarization-maintaining optical fiber before
delivery to the atomic cloud. The probe beams have 1/$e$ diameters
of $3$ mm, $>50$ m radius of curvature at the atom cloud, and are
precisely overlapped ($<$50 $\mu$rad mutual tilt angle) by back
coupling each beam into the opposing beam's fiber launcher. Both
beams are normal to gravity to within 1 mrad. To selectively probe
the $^1S_0$ $(m_J=0)$ to $^3P_1$ $(m'_{J'}=0)$ transition, the
linear probe beam polarization is parallel (within 10 mrad) to a
4.6(2)$\times 10^{-4}$ T bias magnetic field turned on during the
measurement cycle. The counter-propagating beam intensities are
balanced to within 2.5$\%$ by monitoring atomic fluorescence
signals. The single beam intensity (steady-state saturation of
$\sim1.1$) corresponds to a resonant $\pi$-pulse time of $\sim$
87 $\mu$s compared to the probe pulse duration of 100 $\mu$s.
Each point in a given trace of the atomic resonance is
independently normalized by the simultaneously measured atom
number.

Before proceeding further it is important to emphasize that the
counter-propagating beam configuration is essential to obtaining
accurate values for the $^1S_0$ - $^3P_1$ line center frequency.
The ultracold $^{88}$Sr temperature offers an advantage in
reducing systematic frequency shifts associated with atomic
motion. However, we find that due to trapping-beam-intensity
imbalance the atomic cloud acquires typical drift velocities along
the probe beam axis of $\sim$1 mm/s after release from the trap
\cite{werner93}. Although small, this drift corresponds to a
frequency shift of 1.5 kHz if a single probe beam is used for the
spectroscopy. This frequency offset is canceled without perturbing
the line shape if the atomic sample is illuminated by two equal
intensity, counter-propagating probe beams. The presence of both
beams also enables cancellation of the first-order gravity-induced
Doppler shift. For the more realistic $\pm$ 2.5$\%$ intensity
balance achieved here, the line shape is unperturbed but the true
line center remains uncertain by $\pm$ 20 Hz, an effect that could
be eliminated in future measurements by using an optical cavity to
establish the two probe beams. Although the two beam configuration
described here resembles traditional saturated absorption, the
saturation effect is greatly reduced since the excitation pulse
length is comparable to the spontaneous decay time (21 $\mu$s)
while, for the ultra-cold temperatures employed here, the
homogeneous and inhomogeneous linewidths are comparable. We
estimate the saturation effect causes a slight reduction of $\sim$
1$\%$ of the fluorescence signal peak at ($\omega_0$ +
$\omega_R$), where $\omega_R$ is the one photon-recoil frequency
(2$\pi$ 4775 Hz) and $\omega_0$ the original atomic resonance. At
the location of stimulated emission at ($\omega_0$ - $\omega_R$),
the reduction of fluorescence signal is estimated to be $<$
0.1$\%$. Such a small deviation from the originally symmetrical
lineshape causes a systematic shift in the measured resonance
frequency by $\sim$ +4(20) Hz. Note also that Ramsey interrogation
is not well suited to free-space measurements of the $^{88}$Sr
$^1S_0$ - $^3P_1$ transition as its linewidth is slightly larger
than the single-photon recoil frequency. Hence the Ramsey time
required to align the recoil components exceeds the excited state
lifetime.

\begin{figure}
\resizebox{8.5cm}{!}{
\includegraphics{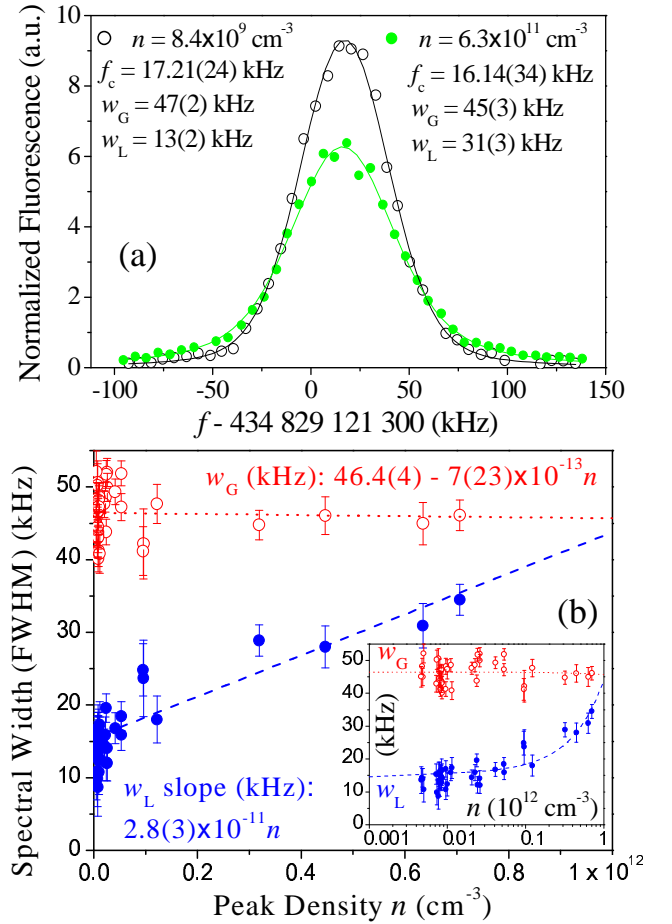}}
\caption{\label{Fig2} (color online) Dependence of the $^1S_0 -
^3P_1$ transition linewidth on $n$ for $10^9$ cm$^{-3} < n <
10^{12}$ cm$^{-3}$. (a) Resonance profiles obtained for two
different $n$. Circles represent data and the solid lines are
fits to Voigt profiles. $f_c$, $w_L$, and $w_G$ are the line
center frequency, the Lorentzian linewidth, and the Gaussian
linewidth, respectively. (b) $w_L$ and $w_G$ vs $n$, indicating a
constant $w_G$ and a linearly broadened $w_L$. The inset is a
semi-log plot of the same data.}
\end{figure}

Figure 2 illustrates the dependence of the $^1S_0$ - $^3P_1$
transition linewidth on the peak density $n$ for $n$ ranging from
$10^9$ cm$^{-3}$ to $10^{12}$ cm$^{-3}$. Figure 2(a) shows two
typical resonance profiles obtained at densities of
$8.4\times10^9$ cm$^{-3}$ and $6.3\times10^{11}$ cm$^{-3}$. The
frequency axis is calibrated with respect to the Cs standard.
Circles represent experimental data while the solid lines are
fits to Voigt profiles. The fits establish the line-center $f_c$,
the Gaussian linewidth $w_G$, and the Lorentzian linewidth $w_L$.
As $n$ increases by a factor of $\sim80$, both $f_c$ and $w_L$
change significantly beyond the associated measurement
uncertainties. As expected, however, $w_G$ remains essentially
constant across the entire density range, as confirmed by the flat
linear fit to the open circles in Fig. 2(b). The measured $w_L$,
shown as filled circles in Fig. 2(b), demonstrates a significant
linear broadening coefficient of $2.8(3)\times 10^{-8}$ Hz
cm$^3$.  The zero-density extrapolated linewidth of 14.5(5) kHz
(Fig. 2(b) inset) agrees with the predicted linewidth of 13.6(5)
kHz after considering power and interaction-time broadenings. In
addition, we find that the integrated area under the Voigt
profile, which is proportional to the number of photons per atom
that contribute to the observed fluorescence signal, remains
constant within the measurement uncertainty for $n < 5 \times
10^{11}$ cm$^{-3}$. For $n \geq 5 \times 10^{11}$ cm$^{-3}$ (peak
optical density of $\sim4.5$), however, the Voigt profile area
decreases steadily with $n$, an effect that arises from probe
beam attenuation and possibly radiation trapping.

\begin{figure}
\resizebox{7 cm}{!}{
\includegraphics{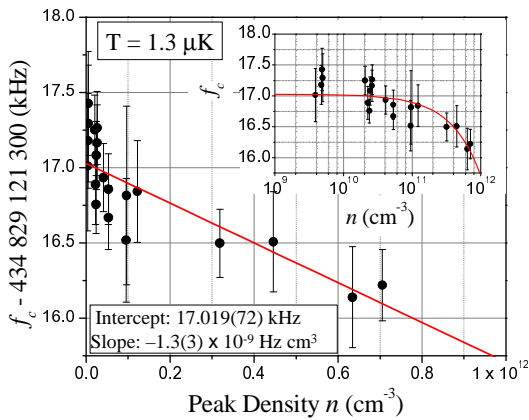}}
\caption{\label{Fig3} Dependence of the $^1S_0 - ^3P_1$
transition line center on $n$. Filled circles (the solid line)
are data (a linear fit). A fractional frequency shift of $-2.9(7)
\times 10^{-24}$ cm$^3$ is determined. The inset is a semi-log
plot of the same data.}
\end{figure}

Figure 3 summarizes the frequency shift of the $^1S_0$ - $^3P_1$
transition vs $n$ for $n < 1 \times 10^{12}$ cm$^{-3}$. Filled
circles (the solid line) are experimental data (a linear fit).
From the fit we determine a fractional frequency shift of $-2.9(7)
\times 10^{-24}$ cm$^3$, $\sim$250 times smaller than the density
shift for Cs \cite{sortais00}. For $n \geq 10^{12}$ cm$^{-3}$
(not shown), observed frequency shifts deviate significantly from
the linear dependence displayed for $n < 5 \times 10^{11}$
cm$^{-3}$. Resolving this deviation requires further
investigation.

The density-related frequency shift and linewidth-broadening
coefficients described above both exceed predicted values based on
general S-matrix calculations of $s$-wave collisions
\cite{julienne04, leo00}. According to Ref. \cite{julienne04}, the
clock shift (in Hz) is $\Delta\nu = 4n(\hbar/m)L_s$, where $m$ is
the atomic mass, 2$\pi$$\hbar$ is Planck's constant, and $L_s$ is
a length that for finite collision temperatures is bound to the
order of $1/k$. Here $\hbar k = \sqrt{mE}$ is the thermal
collision momentum and $E$ is the collision energy. The bound for
the magnitude of the S-matrix is unity. The maximum fractional
frequency shift is thus predicted to be $-0.5 \times 10^{-24}$
cm$^3$, $\sim$$1/6^{th}$ of the measured shift. In the low
temperature limit and in the absence of resonant scattering,
$L_s$ is directly related to the scattering lengths of the two
atomic states associated with the clock transition.

\begin{figure}
\resizebox{6.5cm}{!}{
\includegraphics{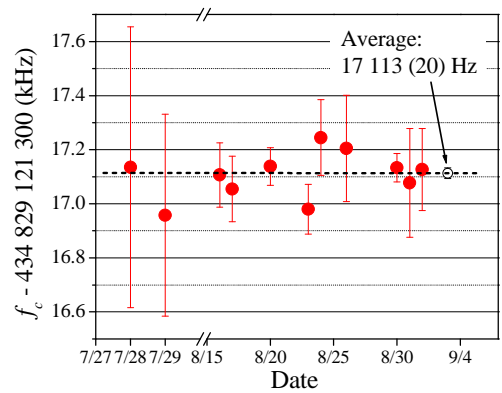}}
\caption{\label{Fig4} Accumulated record of the zero density
$^1S_0 - ^3P_1$ optical clock transition frequency. Each entry is
a weighted mean of several independent measurements that are
referenced to the NIST Cs fountain clock. The overall weighted
mean value is (before correction for systematic shifts) $434\;
829\; 121\; 317\; 113$ Hz, with a statistical uncertainty of 20
Hz.}
\end{figure}

Elastic collisions do not account for the relatively large
linewidth ($w_L$) broadening observed here. For inelastic
contributions, quenching of the $^3P_1$ excited state was
expected to be dominated by $^3P_1$ + $^3P_1$ rather than $^3P_1$
+ $^1S_0$ collisions. The quenching rate constant for $^3P_1$ +
$^3P_1$ collisions has not been determined previously. An
estimate based on the S-matrix formalism described above, however,
limits the linewidth broadening coefficient to $\sim 2 \times
10^{-10}$ Hz cm$^3$, or 140 times smaller than the measured
value. Other mechanisms are then most likely responsible for the
observed linewidth broadening. This conclusion is supported by the
experimental observation that changing the fractional $^3P_1$
density does not significantly alter $w_L$, indicating $^3P_1$ +
$^3P_1$ quenching collisions are not the primary broadening
mechanism. We have also experimentally ruled out any significant
linewidth contributions due to depolarization of $^3P_1$ to the
lower Zeeman level by $^1S_0$ collisions in the presence of the
bias magnetic field. Note, however, that the relatively large Sr
$C_3$ coefficient leads to non-negligible $p$-wave contributions
to the $^3P_1$ + $^1S_0$ process even at a temperature $\sim$400
nK. Thus higher order partial waves may make significant
contributions to the scattering process. Another experimental
observation reveals that for a given range of $n$, a longer
free-flight time (compensated by using a larger initial number of
atoms) gives rise to a larger broadening coefficient. Here it is
likely that the effective thermal collision momentum $k$ among
neighboring atoms is reduced due to their directional
correlations in the free-flight expansion, resulting in a larger
bound for $L_s$ since the bound is set by $1/k$. We note that the
linear fit in Fig. 2 includes the entire data set taken at
various free-flight times (5 ms the shortest time).

\begin{table}
\caption{Systematic corrections and their associated uncertainties
for the absolute frequency of the $^1S_0$ - $^3P_1$ clock
transition.}
\begin{ruledtabular}
\begin{tabular}{lcc}
Contributor & Correction (Hz) & \hspace{0.4mm} Uncertainty (Hz)
\\\hline mutual beam tilt & 0 & 5
\\ position offset & 0 & 2  \\ velocity & 0 & 4 \\ 2nd order Doppler & 0 &
2$\times$10$^{-4}$ \\ 1st order Zeeman & 0 & 0.02 \\ 2nd order
Zeeman & 4 & 0.3 \\ light shift & 0 & 0.01 \\ blackbody shift & 1
& 0.1 \\ recoil shift & -4775 & $<$ 1$\times$10$^{-3}$ \\
stimulated emission dip & -4 & 20 \\ probe power balance
(\textrm{I}) & -5 & 15 \\ probe power balance (\textrm{II}) & 0 &
20
\\ density shift & 0 & 2.9 \\ maser calibration & 0 & 1.7 \\\hline & &
\\ Systematic totals & -4779 & 33 \\
\end{tabular}
\end{ruledtabular}
\end{table}

Figure 4 shows a record of the absolute $^1S_0 - ^3P_1$ optical
frequency accumulated over two months. Each entry represents the
weighted average of multiple independent measurements. A majority
of the measurements are performed in the low density regime
($n<10^{10}$ cm$^{-3}$) and all measurements are corrected for the
density shift shown in Fig. 3. The final weighted mean is $434\;
829 \;121\; 317\; 113$ Hz, with a statistical uncertainty of 20
Hz. A detailed list of the systematic frequency shifts and their
uncertainties are given in Table I. The first four items detail
the residual Doppler effect from the atomic motion in free space.
The next four items describe frequency shifts due to residual
electric and magnetic fields. All have negligible contributions to
the present uncertainty. The photon recoil correction is known to
1 mHz. To ensure that corrections for multiple photon recoils are
unnecessary, we have verified that no significant shift to the
line center occurs for a factor of two decrease in the overall
optical power. Lineshape-asymmetry-related frequency shift caused
by the stimulated-emission photon recoil is $\sim$4(20) Hz. The
other dominant systematic uncertainty arises from the probe beam
intensity imbalance. Along with the 20 Hz uncertainty described
previously (Table 1, probe power balance II), intensity mismatch
leads to preferential multiple scattering in favor of the
stronger beam, resulting in a blue shift due to photon recoil of
5(15) Hz (probe power balance I). The overall correction to the
measured frequency is therefore $-4779$ Hz, with a conservatively
estimated systematic uncertainty of $\sim 33$ Hz. Applying this
correction, the $^1S_0 - ^3P_1$ transition frequency is ($434\;
829\; 121\; 312\; 334 \pm 20_{stat} \pm 33_{sys}$) Hz.

In summary, we have presented the most accurate frequency
measurement of a Sr-based optical frequency standard with a
detailed investigation of possible systematic sources of error.
Using the dense, ultracold sample, we have also performed the
first definitive measurement of density-related frequency shifts
and linewidth broadening for an optical clock transition.

We thank R. J. Jones and T. Parker for help with absolute
frequency measurements and P. Julienne, R. Ciury\l{}o, C. Oates,
and A. Gallagher for useful discussions. Funding is provided by
ONR, NSF, NASA, and NIST.


\begin{references}
\bibitem{hollberg01} L. Hollberg {\it et al.}, IEEE
J. Quant. Electron. {\bf 37}, 1502 (2001).
\bibitem{wilpers02} G.Wilpers {\it et al.}, Phys. Rev. Lett. {\bf 89},
230801 (2002).
\bibitem{katori03} H. Katori {\it et al.}, Phys. Rev. Lett. {\bf 91}, 173005
(2003).
\bibitem{katori99} H. Katori {\it et al.}, Phys. Rev. Lett.
{\bf 82}, 1116 (1999).
\bibitem{binnewies01} T. Binnewies, {\it et al.}, Phys. Rev. Lett. {\bf 87},
123002 (2001).
\bibitem{mukaiyama03} T. Mukaiyama {\it et
al.}, Phys. Rev. Lett. {\bf 90}, 113002 (2003).
\bibitem{loftus04} T. H. Loftus {\it et al.}, Phys. Rev. Lett. {\bf 93}, 073003 (2004).
\bibitem{rafac00} R. J. Rafac {\it et al.}, Phys. Rev.
Lett. {\bf 85}, 2462 (2000).
\bibitem{diddams01} S. A. Diddams {\it et al.}, Science {\bf 293}, 825 (2001).
\bibitem{ye01} J. Ye, L.S. Ma, and J. L. Hall, Phys. Rev. Lett. {\bf 87}, 270801 (2001).
\bibitem{cundiff02} S. T. Cundiff and J. Ye, Rev. Mod. Phys. {\bf 75}, 325 (2003).
\bibitem{ruschewitz98} F. Ruschewitz {\it et al.}, Phys.
Rev. Lett. {\bf 80}, 3173 (1998).
\bibitem{ido03}    T. Ido and H. Katori,Phys. Rev. Lett. {\bf 91}, 053001
(2003). M. Takamoto and H. Katori, Phys. Rev. Lett. {\bf 91},
223001 (2003).
\bibitem{xu03}    X. Y. Xu {\it et al.}, J. Opt. Soc. Am. B-Opt. Phys. {\bf 20},
968 (2003).
\bibitem{park03} C. Y. Park and T. H. Yoon, Phys. Rev. A {\bf 68}, 055401 (2003).
\bibitem{porsev04} S. G. Porsev, A. Derevianko, and E. N. Fortson, Phys. Rev. A {\bf 69}, 021403 (2004).
\bibitem{trebst01}    T. Trebst {\it et al.}, IEEE Trans. Instrum. Meas. {\bf 50}, 535 (2001).
\bibitem{oates04}    C. W. Oates, G. Wilpers, and L. Hollberg, LANL arXiv, physics/0401011
(2004).
\bibitem{courtillot03} I. Courtillot {\it et al.}, Phys. Rev. A {\bf 68}, 030501 (2003).
\bibitem{xu03b} X. Y. Xu {\it et al.}, Phys. Rev.
Lett. {\bf 90}, 193002 (2003).
\bibitem{hong04}    T. Hong {\it et al.}, LANL arXiv, physics/0409051
(2004). R. Santra {\it et al.}, LANL arXiv, physics/0411197
(2004).
\bibitem{ferrari03} G. Ferrari {\it et al.}, Phys.
Rev. Lett. {\bf 91}, 243002 (2003). I. Courtillot {\it et al.},
LANL arXiv/0410108 (2004).
\bibitem{fertig00} C. Fertig and K. Gibble, Phys. Rev. Lett. {\bf 85}, 1622 (2000).
\bibitem{sortais00} Y. Sortais, S. Bize, C. Nicolas, {\it et al.}, IEEE Trans. Ultrason.
Ferroelectr. Freq. Control {\bf 47}, 1093 (2000).
\bibitem{leo01} P. J. Leo {\it et al.}, Phys. Rev. Lett.
{\bf 86}, 3743 (2001).
\bibitem{julienne04}    P. S. Julienne and R. Ciury\l{}o, (private communications, 2004).
\bibitem{hollberg04}    G.Wilpers {\it et al.}, Appl. Phys. B {\bf 76},
149 (2003). U. Sterr {\it et al.}, LANL arXiv, physics/0411094
(2004).
\bibitem{loftus04b} T. H. Loftus {\it et al.}, Phys. Rev. A {\bf 70}, 063413 (2004).
\bibitem{werner93}  J. Werner, {\it et al.} J. Phys. B {\bf 26}, 3063
(1993).
\bibitem{leo00} P. J. Leo, C. J. Williams, and P. S. Julienne, Phys.
Rev. Lett. {\bf 85}, 2721 (2000).
\bibitem{ye03}    J. Ye {\it et al.}, J. Opt. Soc. Am. B-Opt. Phys. {\bf 20}, 1459 (2003).

\end{references}
\end{document}